\documentclass[showpacs,showkeys, amsmath,amssymb, preprint]{revtex4-1}
\usepackage{graphicx}
\usepackage{epsfig}
\usepackage{bm}

\usepackage{fontenc}
\usepackage{graphicx}
\usepackage[dvips]{hyperref}
\usepackage{bm}
\usepackage{amsmath}
\def\beq{\begin{eqnarray}}

\def\eeq{\end{eqnarray}}

\begin{document}

\title{Nonminimal derivative coupling scalar-tensor theories: odd-parity perturbations and black hole stability}

\author{Adolfo Cisterna$^{1}$, Miguel Cruz$^{2}$, T\'erence Delsate$^{3}$ and Joel Saavedra$^{2}$}

\affiliation{$^{1}$Instituto de Ciencias F\'{\i}sicas y Matem\'{a}ticas, Universidad
Austral de Chile, Casilla 567, Valdivia, Chile}
\affiliation{$^{2}$Instituto de F\'\i sica, Pontificia Universidad Cat\'olica de Valpara\'\i so, Casilla 4950, Valpara\'\i so, Chile,}
\affiliation{$^{3}$Theoretical and Mathematical Physics Dept. University of Mons - UMONS\\ 20, Place du Parc - 7000 Mons - Belgium.}

\begin{abstract}
We derive the odd parity perturbation equation in scalar-tensor theories with a non minimal kinetic coupling sector of the general Horndeski theory, where the kinetic term is coupled to the metric and the Einstein tensor. We derive the potential of the perturbation, by identifying a master function and switching to tortoise coordinates. We then prove the mode stability under linear odd-parity perturbations of hairy black holes in this sector of Horndeski theory, when a cosmological constant term in the action is included. Finally, we comment on the existence of slowly rotating black hole solutions in this setup and discuss their implications on the physics of compact objects configurations, such as neutron stars. \footnote{adolfo.cisterna@uach.cl\\
miguel.cruz@ucv.cl \\
terence.delsate@umons.ac.be\\
joel.saavedra@ucv.cl}

\end{abstract}



\maketitle

\date{\today}

\section{Introduction}
Black holes are among the most interesting and fascinating objects in gravitational physics. They form by gravitational collapse of 
heavy dying stars, but the thermodynamical properties they display makes contact to some fundamental aspect, beyond the classic field theory. Indeed, black hole thermodynamics lies between the classical and quantum pictures of nature \cite{JDB1, Haw1, JDB2}. However 
even if black hole thermodynamics represents one of the most important achievements in order to describe gravitational phenomena 
from a quantum field theory point of view \cite{Haw2}, \cite{Haw3}, a complete theory of quantum gravity is still unknown despite the numerous attempts 
in the last decades \cite{Polchinski:1998rr, rovelli, Carlip:2015asa} (and references therein).

Uniqueness theorems are well established for four dimensional asymptotically flat stationary black holes in electro-vacuum. Israel's  theorems adressed the uniqueness of the Schwarzschild and Reissner-Nordstrom black holes \cite{T100}, Wald and Carter extended the result to the rotating case \cite{T101, T102}. In other words four dimensional electro-vacuum stationary black holes in asymptotically flat spacetime  (and regular in the domain of outer communications) are characterised by only $3$ parameters (mass, angular momentum and charge), and are described by the Kerr-Newman metric \cite{T103, T104, T105}. This result seeded the famous phrase `back holes are bald' (they have no hair) \cite{T103}, \cite{T107}.  
\\
Appart from the asymptotically flat case, no definite result were found, except for few cases, for instance for the Kerr (A)dS family.
Given a cosmological constant, this family of solutions depends on two parameters; the mass M and the angular momentum J, which are defined asymptotically as 
conserved charges of the spacetime. These quantities obeys the extremal condition $|J|\le GM^2$, whose saturation gives a 
degenerated horizon, while otherwise the spacetime is nakedly singular. 

Uniqueness theorems are not only important {\it per se}, but, have also inspired the famous Wheeler conjecture ``Black holes 
have no hair" \cite{T108}. This conjecture, established as a theorem in a large variety of cases, is one of the most important and studied 
theorems in black holes physics. Wheeler stated that gravitational collapse leads to a black hole configuration in such a way that 
the final state can only be characterized by mass, charge and angular momentum. Any other additional parameters being regarded as ``hair" are not allowed in the external final description of a black hole.

The no hair conjecture was the arena of the beginning of a frenetic search for hairy black hole solutions. Indeed, after the 
announcement of the conjecture the community started to look for solutions with different contents of matter with the aim of 
obtain some sort of hair. The first hairy solution \cite{volkoviano} was found in the context of Einstein-Yang-Mills theory. 
Many attempt using scalar fields in asymptotically flat spacetimes were done given origin to no hair 
theorems, as is the case for minimally coupled massive scalar fields and proca fields \cite{Bekenstein:1996pn, Bekenstein:1998aw}. 

However, it was recently found in \cite{Herdeiro:2014goa, Herdeiro:2015gia} that hairy black holes in asympotically flat spacetime do exist, in the presence of a complex massive minimally coupled scalar field. In this case, the scalar hair is supported by the rotation, and no static limit exits. See \cite{Herdeiro:2015waa} for a review. Similar conclusions for complex spin-1 fields are suggested in \cite{Brito:2015pxa}, due to the same mechanism than in the scalar case.

Scalar tensor theories have 
been since then a fruitful theoretical laboratory for this respect. 
The first scalar hairy solution was found in \cite{BBM} and independantly in \cite{JDB3} by considering a conformally non minimally coupled scalar field with self interaction, unfortunately 
the scalar field configuration is not regular at the horizon. Adding a cosmological constant this problem can be solved and other families of solutions have been studied \cite{MTZ, dolphi1, Ayon-Beato:2015ada, Anabalon:2012ta}.

The most general scalar-tensor theory with second order equations of motion for the metric tensor and the scalar field has been introduced by Horndeski (dubbed Horndeski theory) \cite{Horndeski} some decades ago. The model has been forgotten untill recently, where it has drawn a lot of attention, especially in the case of cosmology. In fact many inflationary models have been proposed 
\cite{Amendola:1993uh, Saridakis, Suko, Sadjadi:2013psa, Germanis, Germani2, Liender}. The theory has been also investigated deeply in the context of hairy black hole solutions, with special emphasis on the case of non-minimal kinetic coupled scalar fields, particularly in the case where this coupling is given trough the Einstein tensor. Non-hair theorems for the existence of asymptotically flat black hole solutions have been presented \cite{Germani:2011bc, Hui:2012qt}, nevertheless to date, many solutions to this system were found in the case of anti-de Sitter asymptotic behaviors \cite{rinaldi, Kolyvaris:2011fk, charmousis1, dolphi2, Minam1, dolphi3, minas, Zhou} (and references therein). Compact configuration have also been constructed in this sector, leading to constraints on the parameter space of the model \cite{Cisterna:2015yla}.

It was shown in \cite{galileons} that the Horndeski Lagrangian can be written as 
\begin{align}
& L=K(\phi,\rho)-G_{3}(\phi,\rho)\Box\phi+G_{4}(\phi,\rho)R+G_{4,\rho}(\phi,\rho)[(\Box \phi)^2-(\nabla_{\mu}\nabla_{\nu}\phi)^{2}] \nonumber \\
& + G_{5}(\phi,\rho)G_{\mu\nu}\nabla^{\mu}\nabla^{\nu}\phi-\frac{G_{5,\rho}}{6}[(\Box \phi)^3-3\Box \phi(\nabla_{\mu}\nabla_{\nu}\phi)^2+(\nabla_{\mu}\nabla_{\nu}\phi)^3], \label{Lan}
\end{align}
where the functions $K$ and $G_{i}$ depend on the scalar field and its usual kinetic term $\rho=\nabla_{\mu}\phi\nabla^{\mu}\phi$. 
This form of the Horndeski Lagrangian allows us to recognize more clearly important subsectors of the theory, like general relativity, 
Brans-Dicke, the non minimal kinetic coupling sector, etc.
We will mainly focus on the subsector described by the following action 
\begin{equation}
S[g_{\mu\nu},\phi]=\int d^{4}x\sqrt{-g}\left[  \frac{R-2\Lambda}{2\kappa}-\frac{1}{2}\left(\alpha g_{\mu\nu}-\eta G_{\mu\nu} \right)\nabla^{\mu}\phi\nabla^{\nu}\phi \right]. 
\end{equation}
\vspace{0.1cm}

This paper is devoted to derive the perturbation equations for static spherically symmetric spacetimes, with the main purpose of determinate the mechanical stability of black holes in this sector under odd-parity perturbations. 
To this end we shall follow the very general technique displayed in  \cite{Chandra} by Chandrasekhar for the axial perturbations of spherically symmetric spacetimes.
The stability of general relativistic black holes have been studied for a long time, starting with the pioneering works of Reggee and Wheeler \cite{reggewheeler}. Might they have been unstable, their astrophysical importance would probably have been depreciated. Instead they are stable towards linear perturbations. And even more importantly, they are characterised by quasinormal modes (see \cite{qnms} for a review) because of the horizon's absorbant nature. Quasinormal modes are eigen modes of the linear perturbation equations with a complex frequency. The real part of the mode provides a timescale for the perturbation damping, while the imaginary part is the oscillation frequency. In particular, the quasinormal mode ringing describes very well the physics of a binary black hole 'after merger'. 

When considering black hole solutions in a realistic scenario and establishing their possible realizations in nature, the question of stability is essential. For an alternative model of gravity to be viable, configurations known to exist in nature should be well described. In this context, black holes are indeed thought to exist in nature, for instance supermassive black hole seem to play a very important role in galaxies formation, and solar mass black holes are suspected to be present in some binary systems. 

In the context of general scalar-tensor theories with second order equations of motion the group of Kobayashi $et$ $al$. have made an extensive relativistic analysis of linear perturbations. Indeed they have considerer the full Lagrangian (\ref{Lan}) and studied both, the odd \cite{Kob1} and even-parity \cite{Kob2} sectors of perturbations deriving perturbation equations from the second-order actions. Perturbating directly on the action allows them to formulate no-ghost conditions for this kind of solutions. On the other hand in both works \cite{Kob1} and  \cite{Kob2}, the authors also obtain general conditions, necessary but not sufficient, to ensure the gradient stability of static, spherically symmetric solutions.

Note that unstable black holes with a very long unstability timescale might still be relevant in some context \cite{Minamitsuji:2014hha}, otherwise, one could reasonably expect a viable alternative model of gravity to provide stable black holes.

This paper is structured as follows. In section \ref{sec:model}, we detail the sector of Horndeski theory that we are interested in, and present the odd parity perturbation equations.
In section \ref{sec:pert}, we present the parity odd metric perturbation in this model. Next, in section \ref{sec:bh}, we discuss the stability of some black holes solutions in this model.
Finally, we discuss the particular case of the dipolar mode, i.e. the slowly rotating case in section \ref{sec:slowrot} before giving our conclusions in section \ref{sec:ccl}.


\section{Non minimal kinetic coupling sector}
\label{sec:model}
We consider the following action,
\begin{equation}
S[g_{\mu \nu}, \phi] = \int d^{4}x\sqrt{-g}\left[\frac{1}{2\kappa}\left(R - 2\Lambda \right) - \frac{1}{2}\left(\alpha g^{\mu \nu} -
 \beta G^{\mu \nu}\right)\nabla_{\mu}\phi \nabla_{\nu}\phi - V(\phi) \right],
\label{eq:action}
\end{equation}
where $G^{\mu \nu}$ is the Einstein tensor and $\alpha, \beta$ are positive constants to avoid the presence of ghost. The equations 
of motion for the metric $g_{\mu \nu}$ and the scalar field $\phi$ are 
\begin{equation}
E_{\mu \nu} = G_{\mu \nu} + \Lambda g_{\mu \nu} - \kappa \left[\alpha T_{\mu \nu} + \beta\Theta_{\mu \nu} \right] = 0,
\label{eq:eommetric}
\end{equation}
and
\begin{equation}
\nabla_{\mu}\left(\alpha g^{\mu \nu} \nabla_{\nu}\phi - \beta G^{\mu \nu}\nabla_{\nu}\phi \right)  - \frac{dV(\phi)}{d\phi} = 0,
\end{equation}
respectively. In order to write down the equations of motion we have used that $\partial_{\mu}\phi \rightarrow \nabla_{\mu}\phi$, 
the compatibility of the metric $\nabla_{\mu}g^{\mu \nu} = 0$ and the conservation of the Einstein tensor provided by Bianchi 
identity $\nabla_{\mu}G^{\mu \nu} = 0$. For simplicity 
in the notation of equation (\ref{eq:eommetric}), we have defined the following tensors
\begin{eqnarray}
T_{\mu \nu} &=& \nabla_{\mu}\phi \nabla_{\nu}\phi - \frac{1}{2}g_{\mu \nu}\nabla_{\alpha}\phi \nabla^{\alpha}\phi - \frac{1}{\alpha}g_{\mu \nu}V(\phi), \\
\Theta_{\mu \nu} &=& \frac{1}{2}\nabla_{\mu}\phi \nabla_{\nu}\phi R - 2 \nabla_{\alpha}\phi \nabla_{(\mu}\phi R_{\nu)}{}^{\alpha} - \nabla^{\alpha}\phi \nabla^{\beta}\phi R_{\mu \alpha \nu \beta} + 
\frac{1}{2}G_{\mu \nu}\nabla_{\alpha}\phi \nabla^{\alpha}\phi \nonumber \\
&-& (\nabla_{\mu}\nabla^{\alpha}\phi)(\nabla_{\nu}\nabla_{\alpha}\phi) + (\nabla_{\mu}\nabla_{\nu}\phi)\Box \phi \nonumber \\
&+& g_{\mu \nu}\left[-\frac{1}{2}\left(\Box \phi\right)^{2} + \nabla_{\alpha}\phi \nabla_{\beta}\phi R^{\alpha \beta} + \frac{1}{2}(\nabla^{\alpha}\nabla^{\beta}\phi)(\nabla_{\alpha}\nabla_{\beta}\phi)\right].
\end{eqnarray}

\section{Parity odd perturbations}
\label{sec:pert}

In this section we analyze the odd-parity sector of perturbations around spherically symmetric and planar black hole solutions of the action (\ref{eq:action}) in asymptotically (anti-) de Sitter spacetimes. Once we have arrived to our master equation we will analyze the stability of the solutions making use of the Fourier decomposition of the master variable. Then our stability criterium will be based one the positivity of the spectrum of the Schr\"{o}dinger operator, following the lines of \cite{Kodama}.\\
The perturbed metric reads
\begin{equation}
ds^{2} = -A(r)dt^2 + B(r)dr^2 + C(r)\left[\frac{dz^2}{1-kz^2} + (1-kz^2)(d\varphi + k_1 dt + k_2 dr + k_3 dz)^2 \right],
\end{equation}
where $k_1$, $k_2$ and $k_3$ are functions of $(t,r,z)$. $A(r)$, $B(r)$ and $C(r)$ are the metric functions parameterizing the 
most general static background solution of a scalar-tensor theory and the parameter $k = \pm 1, 0$ characterizes the topology
of the spacetime. We consider the following ansatz for the scalar field
\begin{equation}
\phi = \phi_0(r) + \epsilon \Phi(t, r, z),
\end{equation}
where $\phi_0$ is the background field. Considering the Einstein field equations only at first order in $\epsilon$, we find that
\begin{align}
& E^{t}_{r} = \epsilon \kappa \frac{d\phi_0}{dr}\left[\alpha \frac{\partial_{t}\Phi}{A}-\frac{\beta}{ABC}\left(\frac{1}{2}
\frac{A_{r}C_{r}}{A} + \frac{1}{4}\frac{C_{r}^{2}}{C} -kB - C_{r}\partial_{r}\right)\partial_{t}\Phi\right] + \mathcal{O}(\epsilon^{2})= 0, \\
& E^{r}_{z} = -\epsilon \kappa \frac{d\phi_0}{dr}\left[\alpha \frac{\partial_{z}\Phi}{B}-\frac{\beta}{B^{2}C}\left(\frac{1}{2}
\frac{A_{r}C_{r}}{A} + \frac{1}{4}\frac{C_{r}^{2}}{C} - \frac{1}{2}\left(\frac{AC_{r}+CA_{r}}{A}\right)\partial_{r}\right)
\partial_{z}\Phi\right] + \mathcal{O}(\epsilon^{2})= 0,
\end{align}
these equations imply that $\partial_{t}\Phi = 0 = \partial_{z}\Phi$. Substracting the equations $E^{t}_{t}$ and $E^{r}_{r}$, we 
obtain
\begin{align}
& E^{t}_{t}-E_{r}^{r} = \frac{\epsilon \kappa}{B}\left[2\alpha \frac{d\phi_{0}}{dr}\partial_{r}\Phi - \beta \left\lbrace 
\frac{C_{r}^{2}}{BC^{2}}\frac{d\phi_{0}}{dr} - \frac{C_{r}}{BC}\frac{d^{2}\phi_{0}}{dr^{2}} - \frac{2k}{C}\frac{d\phi_{0}}{dr} \nonumber
\right. \right. \\
& \left. \left. + \frac{3}{2}\frac{B_{r}C_{r}}{B^{2}C}\frac{d\phi_{0}}{dr} - \frac{C_{rr}}{BC}\frac{d\phi_{0}}{dr}  - 
\frac{3}{2}\frac{C_{r}A_{r}}{A^{2}BC}\frac{d\phi_{0}}{dr} - \frac{C_{r}}{BC}\frac{d\phi_{0}}{dr}\partial_{r}\right\rbrace \partial_{r}\Phi \right] 
+ \mathcal{O}(\epsilon^{2}) = 0,  
\end{align}
this equation imply that $\partial_{r}\Phi = 0$, using these results in the equation $E^{r}_{r}$ we find
\begin{equation}
E^{r}_{r} = \epsilon \kappa V_{1}\Phi + \mathcal{O}(\epsilon^{2}) = 0, 
\end{equation}
where $V_{1}$ arises from the expansion of the scalar field potential around the background configuration,
\begin{equation} 
 V_{1} = \left. \frac{dV}{d\phi}\right|_{\phi = \phi_{0}}.
\end{equation}
Thus, it follows that $\Phi = 0$, as obtained in the case of a minimally coupled real scalar field \cite{Anabalon:2014lea}. Using 
the zeroth-order equations, it is possible to check that the remaining equations are satisfied up to linear order in $\epsilon$ if 
the following system is satisfied
\begin{align}
& E^{r}_{\varphi} = \frac{\partial}{\partial z}\left[\frac{A}{C}\left\lbrace 1 + 
\frac{\beta \kappa}{2B}\left(\frac{d\phi_{0}}{dr}\right)^{2}\right\rbrace (1-kz^{2})^{2}\left(\partial_{z}k_{2}-
\partial_{r}k_{3}\right)\right]\nonumber \\
& + \frac{\partial}{\partial t}\left[\left\lbrace 1 + 
\frac{\beta \kappa}{2B}\left(\frac{d\phi_{0}}{dr}\right)^{2}\right\rbrace (1-kz^{2})\left(\partial_{r}k_{1}-
\partial_{t}k_{2}\right)\right] = 0,\label{eq:er}\\
& E^{t}_{\varphi} = \frac{\partial}{\partial z}\left[C\sqrt{\frac{B}{A}}\left\lbrace 1 - \frac{\beta \kappa}{2B}\left(\frac{d\phi_{0}}{dr}\right)^{2}\right\rbrace
(1-kz^{2})^{2}(\partial_{z}k_{1} - \partial_{t}k_{3}) \right]\nonumber \\
& + \frac{\partial}{\partial r}\left[\frac{C^{2}}{\sqrt{AB}}\left\lbrace 1 + \frac{\beta \kappa}{2B}\left(\frac{d\phi_{0}}{dr}\right)^{2}
 \right\rbrace (1-kz^{2})(\partial_{r}k_{1} - \partial_{t}k_{2}) \right] = 0,\\
& E^{z}_{\varphi} = \frac{\partial}{\partial r}\left[C\sqrt{\frac{A}{B}}\left\lbrace 1 + \frac{\beta \kappa}{2B}\left(\frac{d\phi_{0}}{dr}\right)^{2}\right\rbrace 
(\partial_{z}k_{2}-\partial_{r}k_{3}) \right]\nonumber \\
& + \frac{\partial}{\partial t}\left[C\sqrt{\frac{B}{A}}\left\lbrace 1 - \frac{\beta \kappa}{2B}\left(\frac{d\phi_{0}}{dr}\right)^{2}\right\rbrace 
(\partial_{t}k_{3}-\partial_{z}k_{1}) \right] = 0\label{eq:ez}.
\end{align}
As we can see, at first order in $\epsilon$, we do not have cosmological constant in the equations of motion. By introducing the 
variable
\begin{equation}
\mathcal{Q} = C\sqrt{\frac{A}{B}}\mathcal{P}_{(+)}(1-kz^{2})^{2}(\partial_{z}k_{2} - \partial_{r}k_{3}), 
\end{equation}
where
\begin{equation}
 \mathcal{P}(r)_{(\pm )} = 1 \pm \frac{\beta \kappa}{2B}\left(\frac{d\phi_{0}}{dr} \right)^{2},
\end{equation}
the equations (\ref{eq:er})-(\ref{eq:ez}) can be written as follows
\begin{eqnarray}
\frac{\sqrt{A}}{C\sqrt{B}(1-kz^{2})^{2}\mathcal{P}_{(-)}}\frac{\partial \mathcal{Q}}{\partial r} &=& -\partial^{2}_{t}k_{3} + 
\partial_{t}\partial_{z}k_{1},\label{eq:1}\\
\frac{\sqrt{AB}}{C^{2}}\frac{1}{(1-kz^{2})\mathcal{P}_{(+)}}\frac{\partial \mathcal{Q}}{\partial z} &=& -\partial_{t}\partial_{r}k_{1} + 
\partial_{t}^{2}k_{2}.\label{eq:2} 
\end{eqnarray}
The combination $\partial_{r}$(\ref{eq:1}) + $\partial_{z}$(\ref{eq:2}) can be written in terms of $\mathcal{Q}$, 
\begin{equation}
\frac{C^{2}}{\sqrt{AB}}\mathcal{P}_{(+)}\frac{\partial}{\partial r}\left[\frac{1}{C}\sqrt{\frac{A}{B}}\frac{1}{\mathcal{P}_{(-)}}
\frac{\partial \mathcal{Q}}{\partial r} \right] + (1-kz^{2})^{2}\frac{\partial}{\partial z}\left[\frac{1}{(1-kz^{2})}\frac{\partial 
\mathcal{Q}}{\partial z} \right] = \frac{C}{A}\partial^{2}_{t}\mathcal{Q}. 
\end{equation}
Using the separation of variables $\mathcal{Q} = Q(r,t)D(z)$, we obtain
\begin{align}
& \frac{C^{2}}{\sqrt{AB}}\mathcal{P}_{(+)}\frac{\partial}{\partial r}\left[\frac{1}{C}\sqrt{\frac{A}{B}}\frac{1}{\mathcal{P}_{(-)}}
\frac{\partial Q}{\partial r} \right] - \gamma Q = \frac{C}{A}\partial^{2}_{t}Q, \label{eq:sep1}\\
& (1-kz^{2})^{2}\frac{\partial}{\partial z}\left[\frac{1}{(1-kz^{2})}\frac{\partial D}{\partial z} \right] = -\gamma D.\label{eq:sep2} 
\end{align}
Let us focus on the case with $k=1$. Setting $z=\cos \theta$ equation (\ref{eq:sep2}) allows to identify $C^{-3/2}_{l+2}(\theta)=D(z)$ 
with a Gegenbauer polynomial with $\gamma = (l-1)(l+2)$ where $l \geq 1$ must be satisfied.
The Gegenbauer polynomials in terms of the Legendre polynomials read
\begin{equation*}
C^{-3/2}_{l+2}(\theta)=\sin^{3}\theta \frac{d}{d\theta}\frac{1}{\sin \theta}\frac{dP_{l}(\theta)}{d\theta}.
\end{equation*}
In this case we introduce the master variable $\Psi(r^{*},t) = \left[C\mathcal{P}_{(-)}\right]^{-1/2}Q(r,t)$ where 
$\frac{\partial}{\partial r} = \sqrt{\frac{B}{A}}\frac{\partial}{\partial r^{*}}$. Using these definitions in equation (\ref{eq:sep1})
one gets the master equation
\begin{align}
& \frac{\partial^{2}\Psi}{\partial r^{*2}} + \left(\frac{1}{2C}\frac{d^{2}C}{dr^{*2}}-\frac{3}{4C^{2}}\left(\frac{dC}{dr^{*}}\right)^{2}
+\frac{1}{2\mathcal{P}_{(-)}}\frac{d^{2}\mathcal{P}_{(-)}}{dr^{*2}}-\frac{3}{4\mathcal{P}_{(-)}^{2}}\left(\frac{d\mathcal{P}_{(-)}}{dr^{*}}
\right)^{2}\nonumber \right.\\
& \left. - \frac{1}{2C\mathcal{P}_{(-)}}\frac{dC}{dr^{*}}\frac{d\mathcal{P}_{(-)}}{dr^{*}} - \gamma \frac{A}{C\mathcal{P}_{(+)}}\right)\Psi = 
\frac{\mathcal{P}_{(-)}}{\mathcal{P}_{(+)}}\partial^{2}_{t}\Psi. 
\end{align}
The mode stability can be explored using the Fourier decomposition of the master variable, $\Psi = \int \Psi_{\omega}e^{i\omega t}dt$,
which yields
\begin{align}
\mathcal{H}\Psi_{\omega} & := -\frac{\partial^{2}\Psi_{\omega}}{\partial r^{*2}} + \left(\gamma \frac{A}{C\mathcal{P}_{(+)}}+\frac{3}{4C^{2}}\left(\frac{dC}{dr^{*}}\right)^{2} - 
\frac{1}{2C}\frac{d^{2}C}{dr^{*2}} +\frac{3}{4\mathcal{P}_{(-)}^{2}}\left(\frac{d\mathcal{P}_{(-)}}{dr^{*}}\right)^{2}\nonumber \right.\\
& \left. - \frac{1}{2\mathcal{P}_{(-)}}\frac{d^{2}\mathcal{P}_{(-)}}{dr^{*2}} + \frac{1}{2C\mathcal{P}_{(-)}}\frac{dC}{dr^{*}}
\frac{d\mathcal{P}_{(-)}}{dr^{*}}\right)\Psi_{\omega} = \omega^{2}_{eff}\Psi_{\omega},\nonumber \\
& = -\frac{\partial^{2}\Psi_{\omega}}{\partial r^{*2}} + V\Psi_{\omega} =  \omega^{2}_{eff}\Psi_{\omega},
\label{eq:H}
\end{align}
where we have defined $\omega^{2}_{eff} = \frac{\mathcal{P}_{(-)}}{\mathcal{P}_{(+)}}\omega^{2}$. It is important to note that even if the scalar field perturbation vanishes, equation (\ref{eq:H}) depends on 
the backreaction produced by the scalar field. If we choose $\beta = 0$, $A = 1-2m/r$ and $C=r^{2}$, this equation becomes the Regge-Wheeler equation. The 
spectrum of the operator $\mathcal{H}$ is positively defined as follows
\begin{equation}
\int dr^{*}(\Psi_{\omega})^{*}\mathcal{H}\Psi_{\omega} = \int dr^{*}\left[\mid D\Psi_{\omega}\mid ^{2} + V_{S}\mid 
\Psi_{\omega}\mid ^{2} \right] - (\Psi_{\omega}D\Psi_{\omega})\mid_{Boundary}, 
\end{equation}
where $D = \frac{\partial}{\partial r^{*}} + S$ and
\begin{equation}
V_{S} = V + \frac{dS}{dr^{*}} - S^{2}.  
\end{equation}
If we choose $S = \frac{1}{2C}\frac{dC}{dr^{*}} + \frac{1}{2\mathcal{P}_{(-)}}\frac{d\mathcal{P}_{(-)}}{dr^{*}}$, we find
\begin{equation}
V_{S} = \gamma \frac{A}{C\mathcal{P}_{(+)}}. 
\end{equation}

\section{Application to black hole solutions}
\label{sec:bh}
For spherically symmetric spacetimes the action (\ref{eq:action}) shows interesting black hole solutions \cite{dolphi1}
\begin{equation}
A(r) = \frac{r^{2}}{L^{2}} + \frac{k}{\alpha}\sqrt{\alpha \beta k}\left(\frac{\alpha + \beta \Lambda}{\alpha - \beta \Lambda}\right)^{2}
\frac{\arctan \left(\frac{\sqrt{\alpha \beta k}}{\beta k}r \right)}{r}-\frac{\mu}{r}+\frac{3\alpha + \beta \Lambda}{\alpha -\beta \Lambda}k, 
\end{equation}

\begin{equation}
B(r)=\frac{\alpha^{2}(\left(  \alpha-\beta\Lambda\right)  r^{2}+2\beta k)^{2}%
}{(\alpha-\beta\Lambda)^{2}(\alpha r^{2}+\beta k)^{2}A(r)}\ , \label{Gsol}%
\end{equation}

\begin{equation}
\left(\frac{d\phi(r)}{dr}\right)^{2} = -\frac{r^{2}\alpha^{2}\left(\alpha + \beta \Lambda \right)\left(\left(\alpha -\beta \Lambda
 \right)r^{2}-2\beta k \right)^{2}}{\kappa\beta \left(\alpha -\beta \Lambda \right)^{2}\left(\alpha r^{2}+\beta k \right)^{3}A(r)},
\label{eq:adolf1} 
\end{equation}
where $\beta \Lambda \neq \alpha$, being $\mu$ a constant, $L^{2} := \alpha/3\beta$ the (A)dS effective radius, $C(r)=r$ and $k = \pm 1$ which characterizes the topology of the spacetime. The case $k=0$ integrate in a different manner given
\begin{equation}
\left(\frac{d\phi(r)}{dr}\right)^{2} = -\frac{\left(\alpha + \beta \Lambda \right)}{\kappa\alpha \beta A(r)},
\label{eq:kcerocase} 
\end{equation}
where $A(r)$ is simply
\begin{equation}
 A(r) = \frac{r^{2}}{L^{2}}-\frac{\mu}{r} =\frac{1}{B(r)}.
\end{equation}
In the expressions (\ref{eq:adolf1}) and (\ref{eq:kcerocase}) we must impose the condition, $\alpha + \beta \Lambda < 0$, in order to have a real scalar field configuration outside the event horizon.

For this solution, we obtain that $\mathcal{P}_{(+)} > 0$, hence because $l \geq 1$, $\gamma > 0$ implies that $V_{S} \geq 0$. Then in any region whenever $A>0$, this implies that all the spherically symmetric four dimensional hairy 
configurations of this family of solutions are mode stable under odd-parity perturbations. In order to be true this statement we must require
\begin{equation}
(\Psi_{\omega}D\Psi_{\omega})\mid_{Boundary} = 0, 
\end{equation}
i.e., vanishing perturbations at the horizon. 



\section{Slowly rotating black holes}
\label{sec:slowrot}
The aim of this section is to attack the existence of slowly rotating black hole solutions. In our analysis this can be archived by setting $k_{2}=k_{3}=0$ and $k_{1}=\omega(r)$ in equation (16). Then the frame dragging function satisfies
\begin{equation}
 \frac{\partial}{\partial r}\left[\frac{C^{2}}{\sqrt{AB}}\left\lbrace 1 + \frac{\beta \kappa}{2B}\left(\frac{d\phi_{0}}{dr}\right)^{2}
 \right\rbrace (1-kz^{2})\frac{\partial}{\partial r}\omega(r)  \right] = 0.
\end{equation}
As concrete example let us take the case where $k=1$ and $z=\cos(\theta)$. Contrary to what happens for minimally coupled scalar fields \cite{Anabalon:2014lea} , the frame dragging equation also depends explicitly on the scalar field configuration and not only through its backreaction. Inserting the solution (30)-(32) we surprisingly obtain for $\omega(r)$
\begin{equation}  
\omega(r)=c_{1}+\frac{c_{2}}{r}
\end{equation}
the same result than in standard Einstein gravity. Note that an interesting and deeper analysis of this results has been recently addressed by Berti $et$ $al$. in \cite{Maselli:2015yva}.
We comment however on the consequences of this result for the physics of compact objects, such as neutron stars. Since the frame dragging equation for compact objects is written around a static configuration, in vacuum, the equation is strictly the same as the one we just presented above. As a consequence, the solution presented here, i.e. the same as in general relativity, is also the \emph{vacuum} solution around a slowly rotating neutron star. As a consequence, this sector of Horndeski theory cannot be distinguished by observing (slowly rotating) neutron stars from a modification of the matter structure. This was discussed in details in \cite{Cisterna:2015yla} for non-rotating solutions, and this results strengthen the conclusions given there.

\section{Conclusions}
\label{sec:ccl}

This work was focused on the analysis of odd-mode stability of black holes in the nonminimal derivative coupling sector of Horndeski theory. This sector is characterized by a non minimal coupling between the kinetic term of the scalar field degree of freedom and the Einstein tensor. For such a coupling and when the dynamics is governed by the linearized equations of motion, we showed that for a real scalar field configuration and spherically symmetric spacetimes the solution described in \cite{dolphi2} is mode stable under odd-parity perturbations. It is important to note that, this can be stated because for our solution $\mathcal{P}_{(+)} > 0$ always holds. This is due to the fact that our scalar configuration is real provided by $\alpha+\beta\Lambda<0$. This reflects the importance of the inclusion of a negative cosmological constant. For instance in \cite{rinaldi} the solution does not contains any cosmological constant term in the action, making the square of the scalar field derivative negative and in consequence $\mathcal{P}_{(+)} > 0$ is not always ensured. Such solution could then exhibits unstable behavior under odd-parity perturbations. In order to obtain our results we have followed the very general treatment given in \cite{Chandra}. With a different approach, in \cite{Kob1}, and without considering explicitly the shape of the solution, conditions on the stability and no-ghost conditions in this kind of solutions have been presented before.
\\
Our result does not depends on the self-interaction term for the scalar field, behaving similar to what has been done for real minimally coupled scalar field in \cite{Anabalon:2014lea}.
\\
The approach developed here allows the construction of slowly rotating solutions. Interesting enough we observe that the frame dragging function behaves exactly like in general relativity. This result have been explored deeply in the context of non hair theorems for galileon slowly rotating black holes in \cite{Maselli:2015yva} and has important consequences in the study of rotating compact configurations such neutron stars. Indeed, this type of solution can be used as the exterior solution of rotating neutron stars in the context of Horndeski theory and in this way search for new constraint on this model. The physical result can be anticipated here from the result we presented: the nonminimal kinetic coupling sector of Horndeski cannot be distinguished from general relativity outside a (slowly) rotating body. More details and applications to neutron stars will be presented elsewhere. 

\section*{Acknowledgments}
The authors are grateful to Andr\'es Anabal\'on for collaboration in early stages of this work. 
A.C would like to thank Julio Oliva for enlightening  comments about boundary conditions. A.C. work is supported by FONDECYT project N\textordmasculine3150157. M.C. work is  supported by PUCV through Proyecto DI Postdoctorado 2015. T. Delsate gratefully acknowledges the FRS - FNRS, as well as the Wallonie - Bruxelles International fund for financial support.

\end{document}